\documentclass{article}

\usepackage{xcolor}
\usepackage{amsmath}
\usepackage{subcaption}
\usepackage{PRIMEarxiv}
\usepackage{booktabs}
\usepackage[utf8]{inputenc} 
\usepackage[T1]{fontenc}    
\usepackage{hyperref}       
\usepackage{url}            
\usepackage{booktabs}       
\usepackage{amsfonts}       
\usepackage{nicefrac}       
\usepackage{microtype}      
\usepackage{lipsum}
\usepackage{fancyhdr}       
\usepackage{graphicx}       
\graphicspath{{media/}}     
\usepackage{tabularx} 
\usepackage{tabu} 
\usepackage{booktabs} 

\title{Robust Melanoma Thickness Prediction via Deep Transfer Learning enhanced by XAI Techniques
}

\author{
  Miguel Nogales \\
  Universidad de Sevilla \\
  Sevilla \\
  \texttt{mnogales@us.es} \\
   \And
  Begoña Acha \\
  Universidad de Sevilla \\
  Sevilla \\
  \texttt{bacha@us.es} \\
     \And
  Fernando Alarcón \\
  C.H.U. Sta. María del Rosell \\
  Cartagena \\
  \texttt{feaso04@gmail.com} \\
     \And
  José Pereyra \\
  Universidad de Sevilla \\
  Sevilla \\
  \texttt{jpereyra@us.es} \\
     \And
  Carmen Serrano \\
  Universidad de Sevilla \\
  Sevilla \\
  \texttt{cserrano@us.es} \\
}

\begin{document}
\maketitle

\begin{abstract}
This study focuses on analyzing dermoscopy images to determine the depth of melanomas, which is a critical factor in diagnosing and treating skin cancer. The Breslow depth, measured from the top of the granular layer to the deepest point of tumor invasion, serves as a crucial parameter for staging melanoma and guiding treatment decisions. This research aims to improve the prediction of the depth of melanoma through the use of machine learning models, specifically deep learning, while also providing an analysis of the possible existance of graduation in the images characteristics which correlates with the depth of the melanomas.
Various datasets, including ISIC and private collections, were used, comprising a total of 1162 images. The datasets were combined and balanced to ensure robust model training. The study utilized pre-trained Convolutional Neural Networks (CNNs).
Results indicated that the models achieved significant improvements over previous methods. Additionally, the study conducted a correlation analysis between model's predictions and actual melanoma thickness, revealing a moderate correlation that improves with higher thickness values.
Explainability methods such as feature visualization through Principal Component Analysis (PCA) demonstrated the capability of deep features to distinguish between different depths of melanoma, providing insight into the data distribution and model behavior.
In summary, this research presents a dual contribution: enhancing the state-of-the-art classification results through advanced training techniques and offering a detailed analysis of the data and model behavior to better understand the relationship between dermoscopy images and melanoma thickness.
\end{abstract}


\section{Introduction} \label{sec:intro}


Melanoma is the most dangerous type of skin cancer\cite{skincancerfoundation2024,aad2024} due to its tendency to spread. It develops in the melanocytes, the cells that produce melanin, the pigment that gives skin its color. Melanoma can spread quickly to other parts of the body and is difficult to treat once it metastasizes. It is potentially deadly if not detected early. Melanomas can develop anywhere on the body and often resemble moles; they are usually black or brown but can also be skin-colored, red, or even blue. Unlike Basal Cell Carcinoma (BCC), melanomas can rapidly change in size, shape, or color\cite{van2015risk,banerjee2000morphological}.

When treating melanoma, the thickness of the tumor, known as the Breslow depth \cite{breslow1970thickness}, is a crucial parameter in determining the stage of the cancer and guiding treatment decisions. The Breslow depth is measured from the top of the granular layer (or from the base of the ulcer, if the cancer is ulcerated) to the deepest point of tumor invasion. It's a strong predictor of prognosis: thicker tumors are associated with a higher risk of metastasis and poorer outcomes. \cite{amouroux2011non}

If melanoma is detected at an early stage, it may be removed with a local excision and a margin of normal skin around it, often leading to a good prognosis. However, if the melanoma has grown thicker or spread to other parts of the body, more extensive surgical intervention might be necessary. This could include wider excision of the skin and underlying tissue, and possibly the removal of nearby lymph nodes to check for the spread of cancer.

Knowing the exact thickness of melanoma is vital for several reasons:

\begin{itemize}
    \item Determining the Surgical Margin: The thickness of the melanoma influences the width of the normal skin margin that needs to be removed along with the tumor. Thinner melanomas require smaller margins, while thicker melanomas require wider margins to ensure all cancerous cells are removed.
    \item Assessing the Risk of Metastasis: Thicker melanomas are more likely to have spread to other parts of the body. Determining the Breslow depth helps in assessing the risk of metastasis and the need for additional treatments, such as chemotherapy, immunotherapy, or targeted therapy.
    \item Guiding Sentinel Lymph Node Biopsy (SLNB): For melanomas of a certain thickness (typically over 0.8 mm, or less in certain high-risk features), a sentinel lymph node biopsy might be recommended to check for cancer spread. Knowing the melanoma's thickness is crucial in deciding whether SLNB is necessary \cite{swetter2021breslowSLNB,Gershenwald2017SLNB}.
    \item Prognostic Value: The thickness of the melanoma is a key factor in staging the disease, which in turn helps in predicting the outcome and survival rate. Thicker melanomas have a worse prognosis compared to thinner ones.
\end{itemize}

In summary, accurately predicting the thickness of melanoma is essential for planning the extent of surgery, assessing the risk of spread, deciding on the need for additional treatments, and estimating the prognosis. Early detection and precise measurement of melanoma thickness are thus critical for improving patient outcomes.

The Breslow depth is recognized as the most valuable prognostic indicator \cite{breslow1970thickness,Ferrandiz}. This index categorizes melanoma into four levels of severity, based on various depth thresholds, which are measured from the granular layer to the deepest malignant cells in the dermis. This ranges are important because they are directly associated with the estimated survival of the patient over the next years. 

\begin{table}[h]
    \centering
        \begin{tabular}{cc}
            \toprule
            Stage & Depth range (mm) \\
            \midrule
            Stage I & 0 to 0.76 \\
            Stage II & 0.76 to 1.5 \\
            Stage III & 1.5 to 4 \\
            Stage IV & More than 4 \\
            \bottomrule \\ 
        \end{tabular}
    \caption{Depth ranges according to the stage of melanoma following the breslow Index}
    \label{table:melanoma_depth}
\end{table}

Another index, which aligns better with the classes of data typically obtained in these cases, is the one created by G. Argenziano et al. \cite{argenziano1999clinical}, which groups melanomas thickness into two groups, based on the one given by the Breslow Index. It basically groups all stages from II to IV into a single group, differentiating between low-depth melanoma ($< 0.76$ mm) and high-depth melanoma ($> 0.76$ mm).

The study of techniques to predict the thickness of the melanoma via image processing is not a sufficiently explored field. In this work, innovative methodologies will try to bridge the gap between traditional dermatological assessment and cutting-edge technology. In \cite{CarvalhoMorphology} Carvalho et al. conducted a dynamic optical coherence tomography (D-OCT) to explore the relationship between the vascular morphology of melanoma and its Breslow index. The research found that melanomas with higher Breslow thickness often exhibit specific vascular patterns, such as dotted and serpiginous branching vessels. These findings suggest a link between vascular characteristics and tumor aggressiveness, enhancing our understanding of the biological behavior of melanoma and its potential for metastasis.

Polesie et al.\cite{Polesie2007} focused on evaluating the precision of international readers, including dermatologists and general practitioners and two machine learning convolutional neural networks (CNN), in measuring the thickness of melanoma from 1056 dermoscopy images. An open Web-based diagnostic study revealed moderate overall accuracy in melanoma thickness assessment, where the collective response of human readers exceeded one of the CNNs but not a pretrained CNN. The findings emphasize the difficulties in accurately assessing the thickness of melanoma and propose that collective human intelligence holds significant value in medical diagnostics. It is also worth mentioning another work by the same main author \cite{Polesie2021}, where a study on the characteristics of melanoma is performed, finding differences and similarities between in situ and invasive melanoma. This is in line with this work, in which invasive melanoma is explored in more depth.

Another contribution to the modeling and prediction of the thickness of melanoma by imagining is given by Saéz et al.\cite{aurora2016}, where instead of Deep Learning-based approaches, machine learning techniques are used, by relying the importance of classification on a set of hand-crafted features. These features are grouped on the basis of color, shape, texture, and pigment network. By using 250 dermatoscopic images of a moderately unbalanced dataset, the model achieves a $0.77$ accuracy and a recall of $0.6$.

Furthermore, Hernández-Rodríguez et al.\cite{hernandez2023prediction}, emphasizes the efficacy of deep transfer learning (DTL) in differentiating between in situ and invasive melanomas according to the Breslow thickness (BT). This finding highlights the potential of DTL as a robust adjunct tool for dermatologists, enhancing the precision of the evaluation of the depth of the melanoma, which is paramount in determining the prognosis and therapeutic strategies. With a dataset of 1315 images, this work achieves a $0.75$ accuracy and $0.58$ F1 score with one of their models while other gets up to $0.65$ recall. They found that DTL approaches can even outperform dermatologists in Breslow Thickness classification.



This work presents a double contribution; first, it seeks to improve the latest state-of-the-art classification results by applying cutting-edge techniques to the train of the neural networks involved, and, on the other hand, it shows a analysis of the image characteristics, seen from two different perspectives. The first one attends to the relations between the images processed by the network, the output of the model, and the other focuses on the distributions of samples analyzing the deep features.

Section \ref{sec:dl} will be devoted to explaining the approach taken to conduct the experiments and exploring the dataset used, all under the Methodology section. In Section \ref{sec:tuning_perf} it will be shown the technical details on how exactly the model was trained and also the resulting metrics of its training. An ablation study is also performed to explain some of the results of using certain methods. Section \ref{sec:feats} will discuss the explanability results obtained from the models. Lastly, conclusions are shown in Section \ref{sec:conclusions}.

\section{Methodology} \label{sec:dl}


In this section, different techniques and processes are discussed. Among them, the dermoscopy database used, the application of neural networks to assess the thickness of the melanoma, and algorithms and data processing. The methodology is then organized into three distinct subsections for clarity. The first subsection is dedicated to the dataset; it describes the characteristics of the dermoscopy images used in this study. The second subsection details the deep learning techniques applied, specifying the architectures of the CNNs and the aspects that most critically affected the training process, as well as the preprocessing applied to the samples. Lastly, the third subsection will go through the explainability tools used to study the features and outputs of the network.

\subsection{Dataset}

In this work, the classification is performed by separating between the classes as in \cite{argenziano1999clinical} and shown in Table \ref{table:melanoma_depth}. Following this approach, Stages II, III, and IV are grouped, obtaining a low-thickness class, composed only of Stage I, and a deeper class with the rest. Arranging the data in these classes is relevant due to two key aspects. The first is the importance of determining whether a melanoma is below or above that particular threshold. As mentioned in Section \ref{sec:intro}, a lession greater than $0.76$ mm is considered a high risk. The second reason is that it also allows for a more balanced dataset, as samples from thicker lessions are more scarce.


\renewcommand{\arraystretch}{1.5} 
\begin{table}[h!]
\centering
\begin{tabular}{|p{3cm}|p{1cm}|p{3cm}|p{1cm}|}
\hline
\multicolumn{2}{|c|}{\textbf{ISIC}} & \multicolumn{2}{c|}{\textbf{Atlas Edra}} \\ \hline
Low Thickness   & 385               & Low Thickness  & 177                \\ 
High Thickness   & 35               & High Thickness   & 89                \\ \hline
\quad \textbf{Total}    & 420               & \quad \textbf{Total}  & 266              \\ \hline
\multicolumn{2}{|c|}{\textbf{Private 1}} & \multicolumn{2}{c|}{\textbf{Private 2}} \\ \hline
Low Thickness  & 168               & Low Thickness  & 107                \\ 
High Thickness    & 119               & High Thickness   & 82                \\ \hline
\quad  \textbf{Total}    & 287               & \quad \textbf{Total}  & 189                \\  \hline
\multicolumn{4}{|c|}{\textbf{Full data set}}  \\
\hline
\multicolumn{3}{|c|}{Low Thickness } & 837 \\
\multicolumn{3}{|c|}{High Thickness} & 325 \\
\hline
\multicolumn{3}{|c|}{\textbf{Total} } & 1162 \\
\hline
\end{tabular}
\vspace{0.4cm}
\caption{Distribution of samples over the subsets}
\label{tab:datasets}
\end{table}

The whole data set is made of a merging of different sources. The first subset is gathered from the ISIC data collection \cite{ISIC}. It is made up of images with information about its thickness, from which a total of $420$ images were obtained. This set is particularly unbalanced. The second one is from the Edras Altas of dermatology \cite{edra}. From this compilation, a total number of $266$ samples is used to contribute to the final set. The other two sets are provided by private institutions, and concieve, respectively, $287$ and $189$ images to the sets. These sets are even more balanced than the Altas set, allowing for a better characterization of the problem. 

A summary of the contribution of each part to the whole data set is shown in Table \ref{tab:datasets} where the amount of samples is divided between the two classes to work with. Regarding the final set, it contains a total of $1162$ samples, which are $837$ from the first class and $325$ from the mix of the other three. It is important to remark the impact of unbalancing in this kind of problems, whereas the ratio between classes is $2.58$. It is also worth to note that from these datasets, the information regarding the biopsy thickness value (not only the breslow class it belongs to) is only preserved in the Private dataset 1 and the Private dataset 2.

\begin{figure}[htp] 
\centering

\begin{subfigure}{0.45\textwidth}
    \centering
    \includegraphics[width=0.9\linewidth]{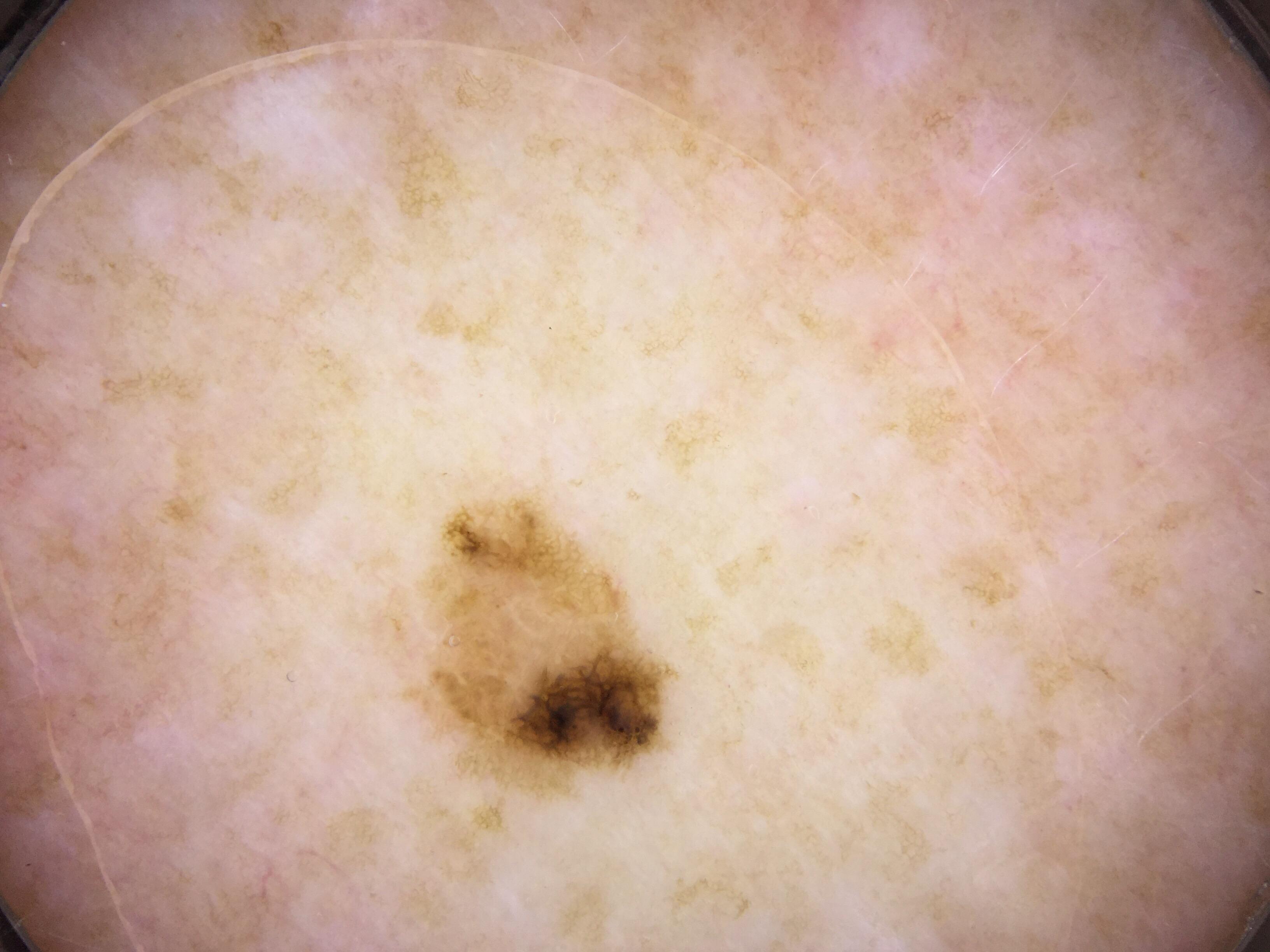}
    \caption{Image from ISIC dataset}
\end{subfigure}%
\begin{subfigure}{0.45\textwidth}
    \centering
    \includegraphics[width=0.9\linewidth]{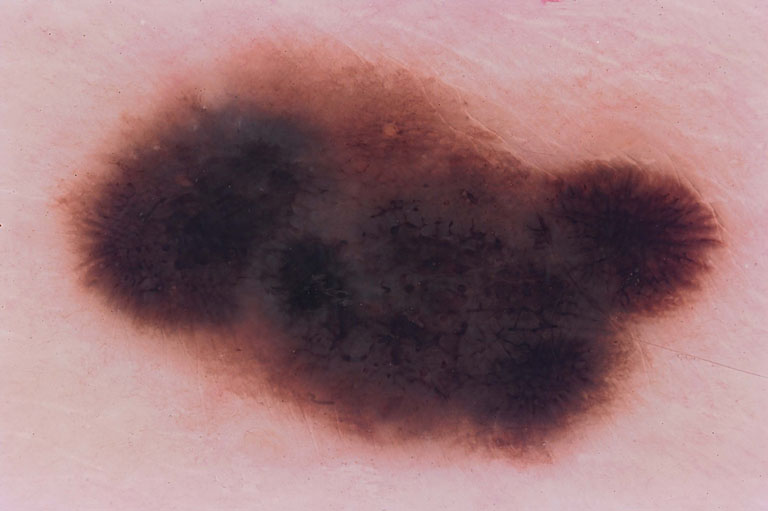}
    \caption{Image from the Edra Atlas of dermtoscopy}
\end{subfigure}

\begin{subfigure}{0.475\textwidth}
    \centering
    \includegraphics[width=0.9\linewidth]{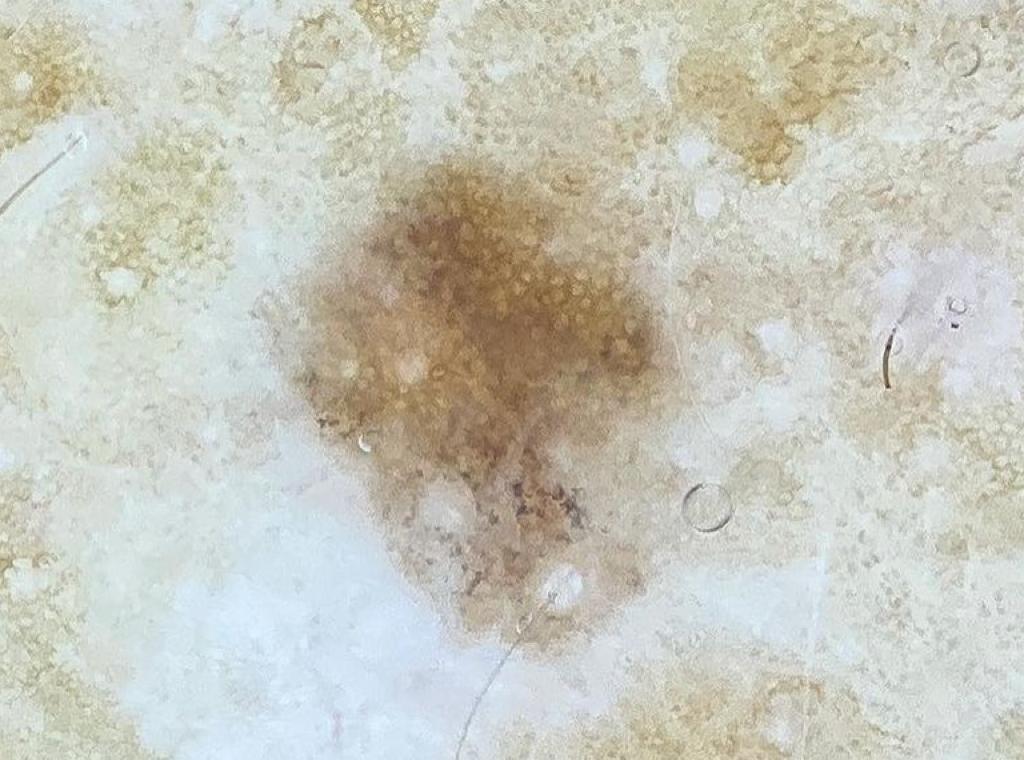}
    \caption{Image from private dataset 1}
\end{subfigure}%
\begin{subfigure}{0.45\textwidth}
    \centering
    \includegraphics[width=0.9\linewidth]{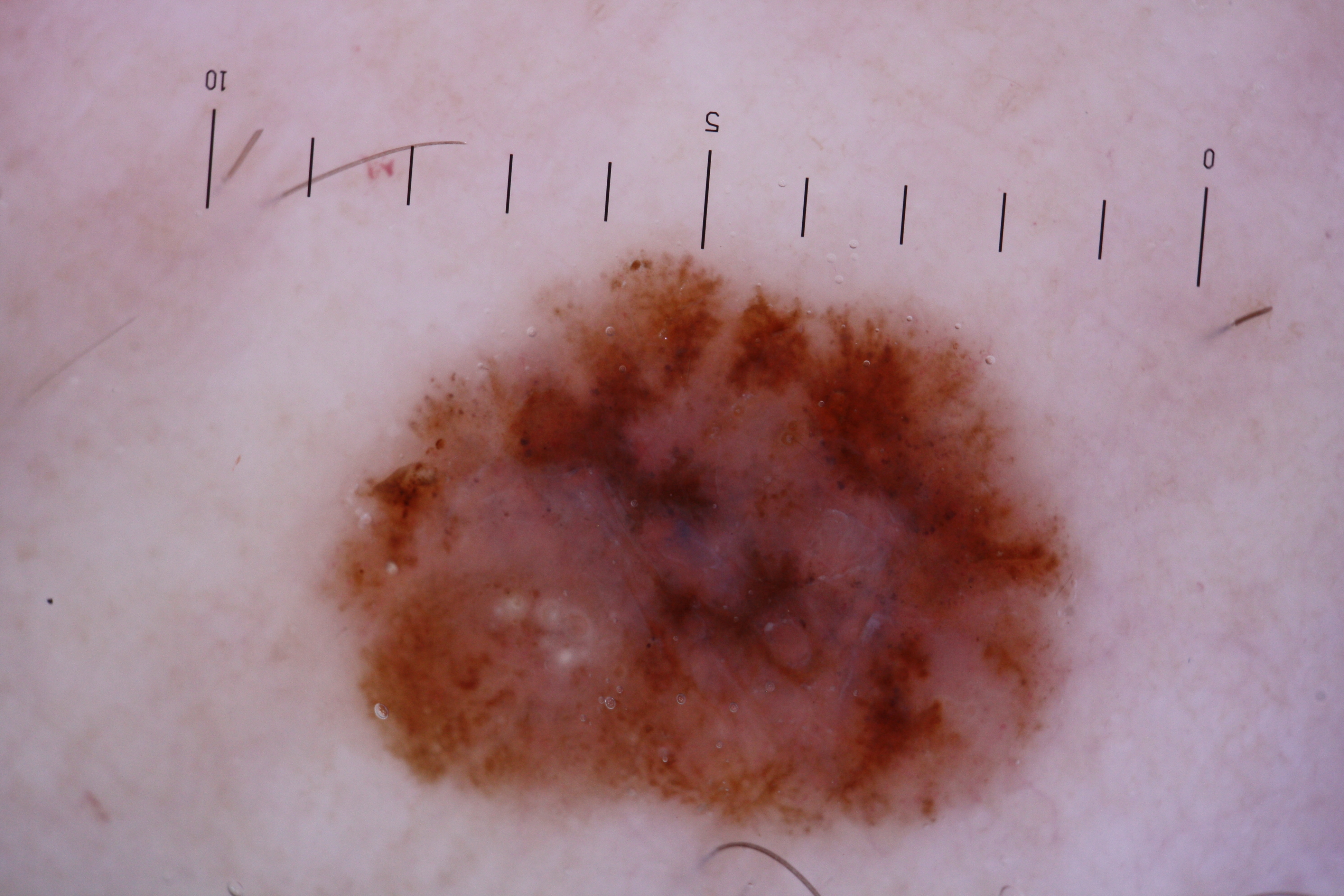}
    \caption{Image from private dataset 2}
\end{subfigure}

\caption{Samples of images from each dataset}
\label{fig:comparision}
\end{figure}

A common problem in the field of Deep Learning, or Machine Learning, is that models tend to focus too much on the given data distribution, which does not represent the whole manifold that exists in the real world. In this work, because images come from different sources, the variability among samples is high, making the database more realistic and resilient. 
Being able to train these types of model with data from such different sources allows for a more robust classification \cite{che2021deep, liu2022deep} with respect to future implementation with real medical systems. In Figure \ref{fig:comparision} a sample of each subset is shown. There are several differences between the images, from straightforward ones such as image definition, shape or color quality to nonvisible ones such as details about when and where the sample was taken, the profile of the specialist who collected the sample, and the instrumentation used.

\subsection{Melanoma Thicknes Prediction}


The first contribution, improving depth estimation of state-of-the-art deep learning models, was achieved through a transfer learning approach. Data were input into a neural network composed of a pre-trained backbone, followed by an additional dense layer. 

\begin{itemize}
    \item \textbf{Feature extraction:} It was performed, as said, with a pretrained backbone with data from the dataset ImageNet \cite{5206848}. Several backbones were tested, finding as most promising CNN's the ones based on ConvNext and RegNet, while Mobilenet v2 being also relevant. Most of the effort spent on tuning hyperparameters was on those configurations. 
    \item \textbf{Image preprocessing:} Resizing, normalization, and cropping were applied to the samples with the default configurations for that particular backbone, so the same preprocessing is used as in the original training. Additionally, on-line Data Agumentation (DA) is used before feeding the samples to the network as a way to improve generalization.  
    \item \textbf{Classifier:} The model chosen to perform the classification is based on linear layers, which varies according to the number of them made up a single dense layer or multilayer perceptron (MLP). Some configurations were tested, ultimely opting to use a simple dense layer.
    \item \textbf{Training:} Training was carried out following a k-Fold cross-validation scheme, supporting the training with enough data and a complete analysis over the entire data set taking it as a test. Training is split into a classifier-only tuning phase, followed by the finetuning of some of the last layers of the backbone alongside the classifier itself. The optimizer was based on Adam for both phases.
\end{itemize}

Key aspects of the training procedure and other relevant choices are explained in more detail in Section \ref{sec:tech_details}. Among the most important techniques used in this work for Breslow depth estimation, whose impact has been shown in the ablation study, three are highlighted. 

\begin{equation}\label{eq:focal}
\text{FocalLoss}(p_t) = -\alpha_t (1 - p_t)^\gamma \log(p_t) \quad \text{with} \quad \gamma = 0 \quad \Rightarrow \quad \text{CrossEntropyLoss}(p_t) = -\alpha_t \log(p_t)
\end{equation}

The first of them is the use of a custom loss function called Focal Loss \cite{lin2018focal_loss}. This function leverages the training by applying weighted loss to samples that are incorrectly classified. This technique has many advantages such as robustness against imbalanced datasets or ease in its hyper-parameter tuning. It is also compatible with class weightening and label smoothing, which are also used. Furthermore, cross-entropy loss can be derived from focal loss when setting the $\gamma$ parameter to zero [\ref{eq:focal}], which weights the importance of a missclassification in the loss. In the equation, the $\alpha$ represents the class weightening weights.

Another relevant addition to the project is the use of the schedule-free optimizer \cite{defazio2024road_schefulefree} based on the Adam optimizer. One technique which is often overlooked when working with transfer learning schemes is the learning rate warm-up and decay to control the gradient descent. This technique adds extra complexity and further extends the hyperparameter space, making tuning more difficult. On the other hand, applying it becomes more relevant as more parameters are optimized, e.g. more pre-trained layers are fine tuned. This new optimizer allows the user to perform this procedure dynamically, adapting the learning rate.
    
The third aspect is the fine-tuning method. It is widely adopted, such as in \cite{Polesie2007, hernandez2023prediction} that the training of the pretrained layers to fine-tune them is performed along with the classifier training, all at once. A more stable and accurate methodology is to perform the classifier training while keeping the whole backbone frozen. When the classifier has reached convergence, a fine-tuning is performed by unfreezing the last convolutional layers or modules of the backbone, with a lower learning rate to account for the increased amount of parameters. This approach is specially important when working with small sized datasets.

The impact of the use of these techniques is analyzed in Section \ref{sec:ablation}, where an ablation study is performed to account for the benefits seen when using all three and its real importance in the final performance of the model.

\subsection{Correlation study and feature analysis}

Understanding and interpreting neural network's results is crucial. Explainable AI enhances trust by making AI decisions transparent, aiding in debugging and improving models, and ensuring accountability in high-stakes applications like this one, assesing melanoma's thickness. Techniques such as feature importance and saliency maps help demystify NNs, ensuring that they operate reliably and fairly. This interpretability not only builds confidence in AI systems but also ensures compliance with regulations, ultimately leading to the responsible and effective use of neural networks.

The second contribution involves an in-depth analysis of the data characteristics, and it is treated as an explanability application on the topic of the melanoma's thickness prediction, helping to support the decisions of the network and also understanding its mistakes. The two techniques employed in this work are:

\begin{itemize}
    \item \textbf{Linear regression analysis:} In this experiment the outputs of the whole network, trained to distiguish between deep and not deep melanomas, are compared against the real thickness value of the lession.     The goal of this experiment is to find if there exists a direct correlation between the output of some model trained to classify the samples between the classes given by the Breslow index and their real thickness. The hypothesis is that there should exist some relationship between the predictions and the samples, in which as the real thickness of the sample grows, the predicted output for the deeper class is higher. In this case, the model would indirectly learn this relationship without the actual label for the real thickness. This would also mean that features representing one class or another are progressively transformed as its thickness becomes more extreme.
    

    \item \textbf{Deep features gradation:} As this features are assumed to be directly related to the depth of the lession, it is of great importance its visualization to confirm the hypothesis made. This visualization is performed in two ways. First by means of Principal Component Analsys (PCA) which creates a set of axis which are ordered by its variance, thus being able to express more information in two dimensions that it would be possible using the deep features directly. Partial Least Squares (PLS) it is also used for this porpuse. This method is similar to PCA but it also involves the labels to perform the subspace formation. By means of this representation a hypotetical gradation would be detected.  
\end{itemize}

These experiments on explanability are intended to find insightful information regarding the data manifold to support the clinical decision-making process.

\section{Model tuning and Performance}\label{sec:tuning_perf}


\subsection{Implementation details}
\label{sec:tech_details}
The data from the detaset mentioned earlier is processed in various ways before being fed into the model. First of all it is normalized, making sure the values are inside the $\{0,1\}$ range. Online data augmentation (DA) is also used in the training phase of the model, providing better generalization. Between the various DA transformations, the ones employed with these models are rotations, zooms, flips, and addition of Gaussian blur. No DA technique with respect to color is used in any way to ensure robustness. All images are resized beforehand to $256$x$256$ pixels.

The underlying architectures for the models (i.e. the neural network backbone) have been pre-trained with the Imagenet dataset \cite{5206848}, and are used following a deep transfer learning approach. Although several architectures have been tested in this work, the most successful one have been Regnet \cite{radosavovic2020designing} and ConvNext Tiny\cite{liu2022convnet_ConvNext}. 

To assess the performance of the model, a stratified K-fold cross-validation approach is employed. This technique ensures accurate model evaluation in cases of limited data and addresses imbalances, such as in this case. The selected $k$ is $10$, resulting in a fair balance between the time spent training and the quality of the validation. 

Two main regularization techniques are used to prevent overfitting of the models. Firstly, dropout, which is a staple in neural network regularization which has been applied with a ration of $0.3$. As a different way of regularization, the loss function used, called focal loss, also acts as a regularization method by weighting samples by their classification complexity given the model as seen in \ref{eq:focal}, with $\gamma$ set to $0.3$.

Class weighting addresses the imbalance further by assigning more weight to underrepresented classes, improving the model's performance on these classes and denying the scenario where the model only predicts the mayoritary class to improve the loss. Instead of using the inverse of the frecuency, which is a typical approach, the weights were hardcoded so the weight of the second class is five times bigger than the not deep class. 

The use of focal loss is strategic for focusing on hard-to-classify instances, dynamically adjusting the cost function to prioritize misclassified samples, which is beneficial for handling class imbalances. The Adam optimizer is chosen for its efficiency in handling sparse gradients and its adaptability to large datasets and high-dimensional spaces, used in conjunction with the scheduler-free scheme proposed in \cite{defazio2024road_schefulefree}.

The training of the model is performed in stages, starting with the head and gradually fine-tuning deeper layers, allowing for a more controlled and effective learning process. Initially, the model adapts to high-level features using pre-trained weights, and then it refines its understanding based on the specific task. Dropout prevents overfitting by randomly omitting units from the network during training, compelling the model to learn more generalizable features. The learning rate for the first phase of the training was $10^{-3}$, for 50 epochs, while the fine-tuning lasted the same number of epochs with a learning rate of $10^{-5}$. Four convolutional blocks were fine-tuned in the case of ConvNext.

The batch normalization layers are not retrained, which helps stabilize the learning process and speed up the training. This approach leverages the normalization parameters learned from the pre-trained network, avoiding unnecessary adjustments during fine-tuning. 

Regarding the models, several pre-trained weights shared by PyTorch for some architectures have been tested, highlighting the Regnet and ConvNext families, performing better than the rest.

Lastly, the classifier used with these pre-trained backbones is just a dense layer, preceded by a Global Average Pooling layer after the last convolutional layer, with two outputs, one for each class. Do note, only the feature extractor of these models has been used with a TL approach, the classifier is trained from zero.

\subsection{Results}

\begin{table}[ht]
\centering
\begin{tabular}{@{}lcccc@{}}
\toprule
Model         & Recall      & Precision       & Accuracy       & F1         \\ \midrule
ResNetV2 \cite{hernandez2023prediction} & 0.60  & 0.57 & 0.70  & 0.57  \\ 
InceptionV3 \cite{hernandez2023prediction} & 0.65  & 0.53 & 0.65  & 0.54   \\ 
EfficientNetB6 \cite{hernandez2023prediction} & 0.58  & 0.61 & 0.75  & 0.58   \\ 
RegNet & 0.74 & 0.60 & 0.78 & 0.66 \\
ConvNext & 0.79 & 0.60 & 0.79 & 0.67 \\
\bottomrule
\\
\end{tabular}
\caption{Performance metrics of the models}
\label{tab:results}
\end{table}

 The results against the benchmark model are shown in Table~\ref{tab:results}. The results obtained from our models represent a significant improvement over previous attempts using this transfer-learning approach. It has to be noted that, as data is scarce, databases from these two different works are different, but they feature a similar amount of samples (1315 vs. 1162) and two sources, the ISIC and one of the private sets, are the same. There are also differences in the unbalancement, as in this work there is a higher skewness. Furthermore, the data set used in this work is composed of four different sources with a similar contribution from each of them, as said, helping with robustness. Sources for the reference work come from three different origins, with one of them representing more than three quarters of the total size. The reason behind the use of those models as comparative is due to several factors such as good practices, the size of the dataset similar to ours, transparency, and currency.

Table \ref{tab:results} summarizes the metrics obtained and compares them between models. A slight improvement in accuracy and F1 can be seen, while the recall improvement is more noticeable. It is also worth remarking that all metrics are improved, avoiding the maximization of one particular metric to the detriment of the other ones. ConvNext is the top-perfoming backbone, which may be due to the mix between the inherent performance of the CNN with its reduced amount of parameters when using the tiny size. The number of deep features is also smaller than that of other popular backbones, which might help with the convergence of the dense classifier.

The performance enhancement can be attributed to the meticulous application of various training techniques. Techniques such as k-fold cross-validation, online data augmentation, strategic class weighting, and the introduction of focal loss have played a pivotal role in optimizing the learning process. Furthermore, the adoption of advanced schedule-free optimizers like Adam and the use of dropout have contributed to a more robust and generalizable model. These methodological refinements have led to a notable enhancement in the model's predictive accuracy and overall reliability.
 
Furthermore, in the context of classifying high-depth melanoma using machine learning models, achieving a high recall rate is critically important. Recall, or sensitivity, measures the ability of the model to correctly identify all positive cases, which in this scenario are the actual instances of high-depth melanoma. Given the serious implications of missing a high-risk melanoma case, prioritizing recall helps ensure that fewer cases go undetected. A model with high recall may occasionally misclassify benign conditions as malignant, but this trade-off is preferable to missing potential melanomas, which can be life-threatening. Therefore, optimizing for high recall minimizes the risk of false negatives, thereby increasing the likelihood that patients with serious conditions receive the necessary medical attention immediately. This approach aligns with the primary goal of medical diagnostics: to maximize patient safety and treatment efficacy. For this reason, special importance has been given to improving recall, always without compromising the other metrics.
 
\subsection{Ablation study} \label{sec:ablation}

To assess the importance of the methods used, an ablation study was carried out. The elements to study in this section will be the loss function, the focal loss, the optimizer, and the fine tuning method. All these tests are performed with the ConvNext backbone.


The first, focal loss, aims to apply a higher weight to samples that have been incorrectly classified, acting as a kind of reinforcement. To perform this test, the parameter $\gamma$ of the focal loss is set to zero, which makes it equal to the cross-entropy loss, as previously shown in Equation \ref{eq:focal}. The results can be seen in Table \ref{tab:ablation}, where a significant drop in performance is observed. This ocurrence is likely due to the fact that without the focal loss giving more significance to missclasified samples, the model tends to the most represented class. In this case, it is the one with the least samples. This happens because as the class weights used with this configuration are five times bigger for the underrepresented class, its impact on the loss function is greater.

\begin{table}[ht] 
\centering
\begin{tabular}{@{}lcccc@{}}
\toprule
         & Recall      & Precision       & Accuracy       & F1         \\ \midrule
ConvNext  & 0.79  & 0.60 & 0.79  & 0.67  \\ 
No focal loss & 0.99  & 0.30 & 0.36  & 0.46  \\ 
No scheduler  & 0.75  & 0.60 & 0.79  & 0.67  \\ 
Single phase FT & 0.76  & 0.56 & 0.76  & 0.64  \\ 

\bottomrule
\\
\end{tabular}
\caption{Ablation study metrics}
\label{tab:ablation}
\end{table}

With respect to the optimizer, it can be seen that it does not affect the performance too much compared to the loss function, apart from a slight drop in recall. This may be due to the small size of the data set, related to the number of parameters updated in each iteration. However, it helps with convergence, making descent to the optimum smoother, enabling higher learning rates.

Lastly, instead of making the two-phase training, composed of first a training of only the head and then a fine tuning of the deeper layers of the pre-trained backbone, the training of both parts is done at the same time. This is useful for convergence speed, but it makes the models perform worse; in this case, the overall performance is worsened a little bit.

\section{Study on the melanomas characteristics} \label{sec:feats}

First, this work will focus on the network output, developing relationships between the real world cases and the training cases of the network, taking advantage of the real value for the thickness of the lession. The second analysis will assess the relationships between the features of the network and their representation with visualization techniques, projected in the PCA space. With these tools, we expect to shed more light on the study of the thickness via dermatoscopic images.

\subsection{Correlation analysis of the networks output}


The training was carried out considering two classes, as explained in Section \ref{sec:dl}, where stages other than the first were merged. A subset of the images kept their biopsy thickness value, and when those images were selected as a test in the k-fold split, the network outputs for those were collected. It is worth noting that the network is the same as the one shown in the previous section. Using k-fold cross validation allows the use of all the images as test in some point. This is particulary relevant due to the fact that not all the images in the dataset have their biopsy thickness values, then helping to make a large enough test set.


\begin{figure}[h!]
  \centering
  \includegraphics[width=0.9\textwidth, trim=0 0 30 0, clip]{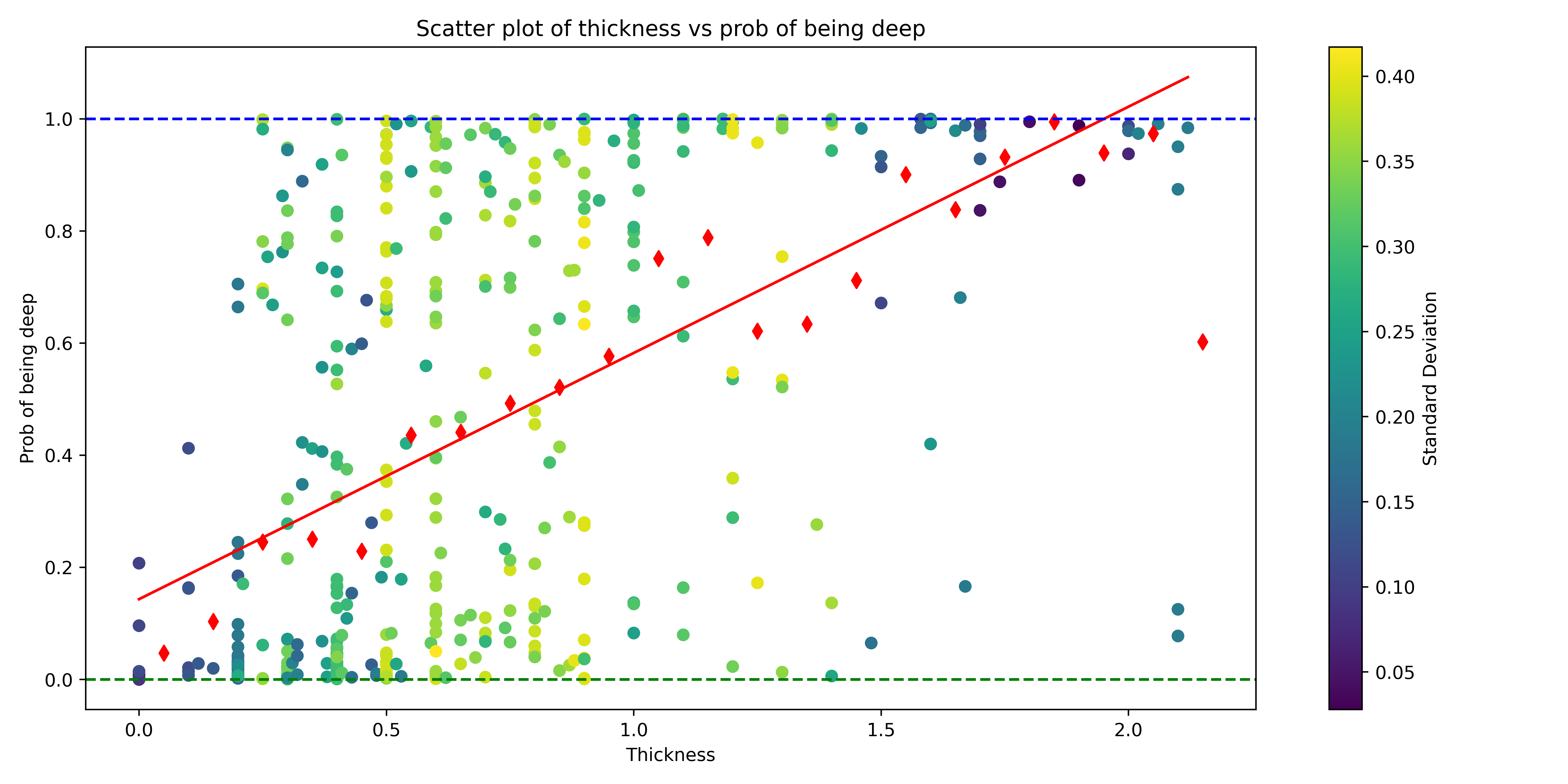}
  \caption{Regression of the deepness based on output prediction.}
  \label{fig:reg_pred}
\end{figure}

\begin{table}[t]
\centering
\begin{tabular}{l|c}
\hline
Ranges [mm] & \( R^2 \) Value \\
\hline
0 to 0.4 & 0.38 \\
0.4 to 1 & 0.08 \\
more than 1 & 0.44 \\
\hline
\end{tabular}
\vspace{0.3cm}
\caption{$R^2$ values for segments.}

\label{tab:r2_values}
\end{table}

Following this idea, from the outputs of the network, a simple linear regressor along with second- and third-order regressors were trained to fit the points. The coefficients $R^2$ for these curves are low, a $0.25$ value for linear regression, which did not improve for the higher-order ones, which stayed at $0.26$. The regression plot is shown in Figure \ref{fig:reg_pred}. The red line represents the linear regression line, with red markers representing the mean of the points in the $0.1$mm interval where they are placed. Data samples are also presented in different colors depending on the standard deviation of that group, showing an interval in the middle where the data is more dispersed.

The output of the model has a very high variation, classifying samples as one of each classes with high confidence when they are not, specially around certain point. This interval can be seen approximately from $0.4$mm to $1$mm, where the data is specially spread. This suggests that the samples in that range could have mixed features from both deep and not deep thicknesses.  To quantify this occurrence, the $R^2$ value for the ranges studied is calculated and shown in Table \ref{tab:r2_values}, ensuring that there is a large difference between segments. This can also be seen in the plain scatter plot, as data in that range are predicted to be as deep with any probability.




\begin{figure}[t!]
  \centering
  \includegraphics[width=0.9\textwidth]{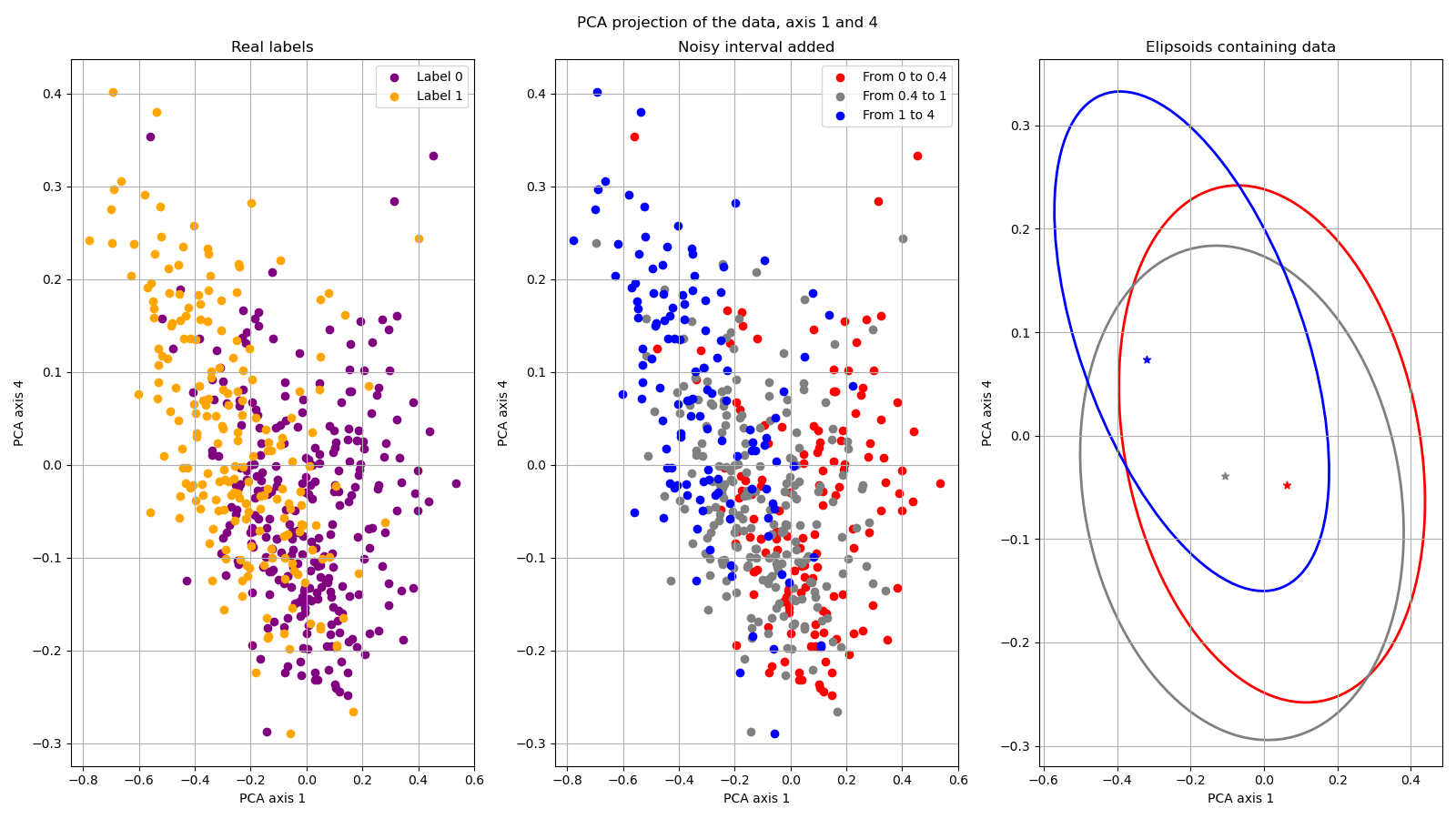}
  \caption{PCA projection applied to the deep features.}
  \label{fig:pca}
\end{figure}

\begin{figure}[t!]
  \centering
  \includegraphics[width=0.9\textwidth]{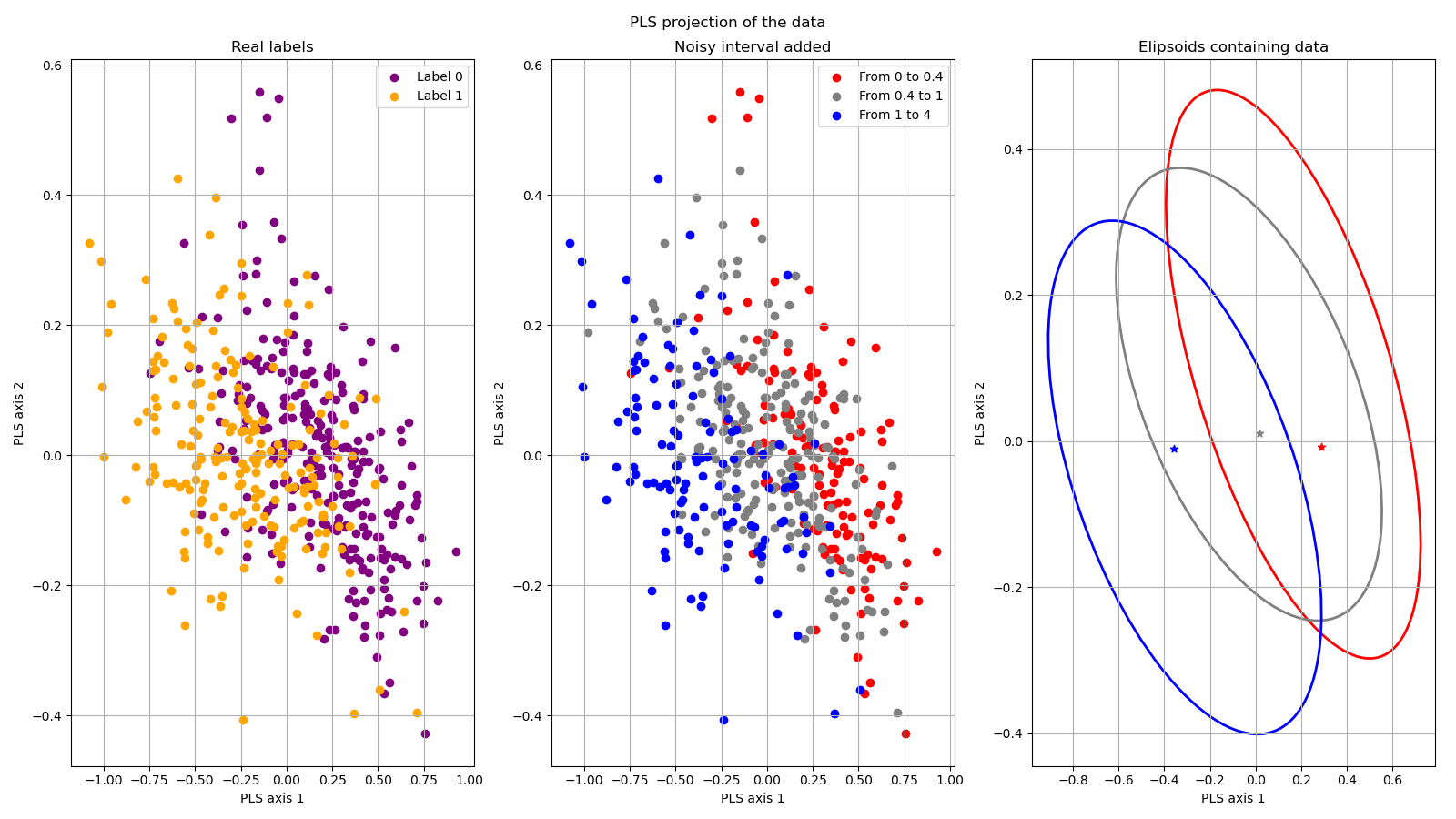}
  \caption{PLS projection applied to the deep features.}
  \label{fig:pls}
\end{figure}

\subsection{Deep features dimension visualization}

Recognizing that a subset of the samples seems to have mixed characteristics, an exploration of the deep features is performed. These features were projected into two subspaces in order to analyze how the distribution of samples would manifest. This is seen as a way to explain the behavior of the data and the model, as the deep features contain the information used to understand the images.

\begin{figure}[t!]
  \centering
  \includegraphics[width=0.9\textwidth]{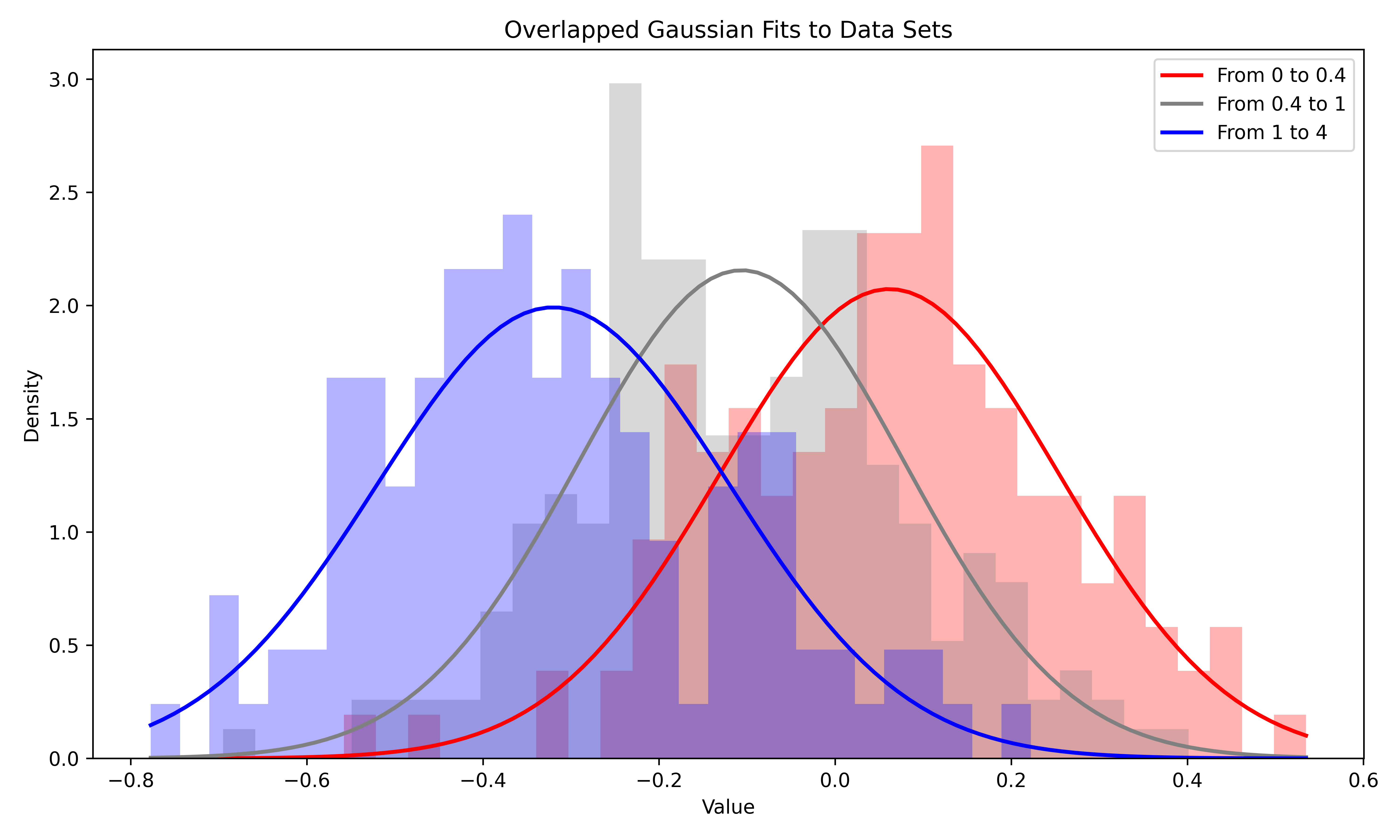}
  \caption{Gaussians fitted to data in one dimension of PCA basis.}
  \label{fig:1d_gauss}
\end{figure}

The first subspace is obtained by choosing the most relevant PCA axes. This method provides a set of basis ordered by variance with respect to the samples. For visualization purposes, only the first two more relevant axes are selected. It should be noted that relevant in this case does not account for the axes with higher variance, but for those that make the interclass variance higher. In this case, those axes are the first one and the fourth. The second approach consisted of projecting the set over the axes given by a PLS subset. The reason behind these choices is to have two different bases to project the features into, in the case of PCA a general basis, which is not influenced by the labels of the models, just the features. The case of PLS is the opposite, where the labels are involved, producing a different set of axes.


Figures \ref{fig:pca} and \ref{fig:pls} show the distribution of the samples on those PCA and PLS axes. The first and second subplot show the data projected into the axes of the basis created by its respective method with the higher variance. In the first, the colors represent the original labels, setting $0.76$ mm as the threshold between them. In the second one, the interest interval is shown in gray, revealing the noisy characteristics of this group. Lastly, in the third subplot, ellipses fitted to the points are shown as a way to visualize the position of the distributions as well as their means. It can be clearly seen how the deep and not deep classes are relatively separated, which is not the case for the middle range, as those samples seem to be more scattered. The mean of the distributions keeps the slight gradation by its thickness studied in the previous subsection, but the variance, easier to visualize in the ellipses plot, makes the distrubitions overlap. It is relevant how in PCA the gray ellipse covers a great part of the other two, supporting the hypothesis that samples in that interval have features from the other two. This does not happen with PLS, likely due to PLS using the labels themselves to perform the regression. 

Lastly, Figure \ref{fig:1d_gauss} shows the same distribution as the one seen in the PCA projection but expressed only on the axis with the highest variance. Gaussians have been fitted to the distributions of samples to show how they relate to the data, which is shown at the back with lower opacity. The extremes groups, with the higher and lower thicknesses, fit closely to the distribution, while the middle one does not match that much.

\section{Conclusions} \label{sec:conclusions}

This study demonstrated the effectiveness of applying advanced machine learning techniques, specifically deep learning, to enhance the accuracy of predicting melanoma thickness using dermoscopy images. Through a robust methodology involving pre-trained CNNs, it has been addressed the challenge of accurately determining the Breslow depth, a critical factor in melanoma staging and treatment planning, improving state-of-the-art results with a robust validation scheme.

Our models achieved significant improvements in recall, which is pivotal in medical diagnostics, where high recall reduces the risk of false negatives, a critical outcome for the detection of melanoma. The integration of rigorous validation methods such as k-fold cross-validation contributed to these advancements and to the veracity of the results shown, added to the inherent diversity found in the dataset, contributing to the robustness of the network.

The study also explored the correlation between the computed features from the deep learning models and actual melanoma thickness, revealing insights into how machine learning can approximate complex biological and medical relationships. Although the correlation analysis indicated moderate relationships, it emphasized the potential of machine learning in enhancing our understanding of the characteristics of melanoma beyond traditional imaging techniques.

Furthermore, the use of explainable AI techniques provided deeper insights into the decision-making process of neural networks, enabling a more transparent, reliable, and interpretable approach to medical diagnostics. The application of feature importance and dimensionality reduction techniques such as PCA and PLS facilitated the exploration of the underlying patterns in the data, supporting the clinical decision-making process.

For future work, it would be beneficial to expand the datasets to include a wider range of melanoma types and stages from diverse demographic backgrounds to further enhance the generalizability of the models. Further research could also explore the relation between deep features and dermatoscopic features, which dermatologists use. Another approach for research could be refining the predictive capabilities of the models or other types of images, such as thermal ones. In addition, real-time diagnostic systems could be developed to assist dermatologists in clinical settings, providing immediate insights and recommendations based on AI-driven analyses. It would also be beneficial to continue with the explanability setting, being able to obtain more information into the network decisions, for example, by means of explaining each individual deep feature and checking the more relevant ones.

\section*{Acknowledgments}
This work was supported by the Andalusian Regional Government (PROYEXCEL\_00889) and MCIN/AEI/10.13039/501100011033 (Grant PID2021-127871OB-I00), and ERDF/European Union (NextGenerationEU/PRTR).

\bibliographystyle{unsrt}  
\bibliography{references}

\end{document}